\documentclass[aps,pra,twocolumn,showpacs]{revtex4}
\usepackage{amsmath}
\usepackage{epsfig}
\usepackage{amssymb}
\usepackage{graphicx}

\begin{document}
\date{\today}

\title{Stability of excited states of a Bose-Einstein condensate in an anharmonic trap}

\author{Dmitry A. Zezyulin}
\email{d.zezyulin@gmail.com}

\author{Georgy L. Alfimov}
\email{galfimov@yahoo.com}

\affiliation{ Moscow Institute of  Electronic Engineering,
Zelenograd, Moscow, 124498, Russia.}

\author{Vladimir V. Konotop}
\email{konotop@cii.fc.ul.pt}

\affiliation{Centro de F\'{\i}sica  Te\'{o}rica e Computacional, Universidade de
Lisboa,
 Complexo Interdisciplinar,  Avenida Professor Gama Pinto 2, Lisboa
1649-003, Portugal and Departamento de F\'{\i}sica, Faculdade de
Ci\^encias, Universidade de Lisboa, Campo Grande, Edif\'{\i}cio C8,
Piso 6, Lisboa 1749-016, Portugal}

\author{V\'{\i}ctor  M. P\'erez-Garc\'{\i}a}
\email{victor.perezgarcia@uclm.es}
\affiliation{Departamento de Matem\'aticas, E. T. S. I. Industriales,  and
Instituto de Matem\'atica Aplicada a la Ciencia y la Ingenier\'{\i}a,
Universidad de Castilla-La Mancha, 13071 Ciudad Real, Spain.}

\begin{abstract}
We analyze the stability of non-ground nonlinear states of a Bose-Einstein condensate
 in the mean field limit in
effectively 1D (``cigar-shape'') traps for various types of
confining potentials. We find that nonlinear states become, in
general, more stable when switching from a harmonic potential to
an anharmonic one.  We discuss the relation between this fact and the
specifics of  the harmonic potential which has an equidistant spectrum.
\end{abstract}


\maketitle

\section{Introduction}\label{intro}

The mean field theory of a Bose-Einstein condensate
(BEC) is based on the  Gross-Pitaevskii equation (GPE) \cite{Pitaev}
\begin{equation}
i \Psi_t=-\Delta\Psi+V({\bf x})\Psi-\sigma |\Psi|^2\Psi ,
\label{GPEq}
\end{equation}
for the macroscopic wave function $\Psi=\Psi(t,{\bf x})$. This model
describes accurately the behavior of an ultracold condensed atomic
cloud trapped by an external potential $V({\bf x})$. In the
dimensionless variables of Eq. (\ref{GPEq}) the Planck constant is
equal to one, $\hbar=1$, and the atomic mass is $m=1/2$.
The parameter  $\sigma$ stands for the sign opposite to  that of the
scattering length $a_s$: $\sigma=-$sign$\,(a_s)$.
An important class of solutions of GPE are
stationary nonlinear modes defined as
\begin{eqnarray}
\Psi(t,{\bf x})=e^{-i\omega t}\psi({\bf x}),
\label{Anzats}
\end{eqnarray}
with the boundary conditions
\begin{eqnarray}
\psi({\bf x})\to 0\quad \mbox{as}\quad |{\bf x}|\to
\infty.\label{Local}
\end{eqnarray}
In this case $\psi({\bf x})$ satisfies the equation
\begin{equation}
\Delta\psi+(\omega-V({\bf x}))\psi+\sigma \psi^3 =0.\label{StatGPEq}
\end{equation}
The nonlinear eigenvalue $\omega$ is called  the chemical
potential in the context of BEC applications. The {\it ground
state} solutions of Eqs. (\ref{Local})-(\ref{StatGPEq}) (i.e. its
positive solutions which minimize the energy functional for Eq.
(\ref{GPEq})) are of primary importance for BEC applications
\cite{EdwBur1995}.
Apart from them some {\it non-ground} nonlinear modes have also been
studied (see e.g.
\cite{Yukalov1997,Kivshar2001,MalomLasPhys2002,AgPres2002,KonKev,KevrekDyn}).
However all of the practical applications of high order modes are linked to their experimental
feasibility, what requires stability. The stability
of high order modes has already been studied in the case of a harmonic
potential $V({\bf x})=|{\bf x}|^2$. The one-dimensional case was studied in Refs. \cite{Carr2001,KevrekDyn,AlfZez2007,PelKevr2007} while multidimensional solutions were considered in Refs.
 \cite{VP1,VP2,VP3,VP4,Fin1,Michalache2006,Fin2,RadSymmKevrek2007,Watanabe}.

However, to the best of our knowledge, the relation between
the  stability properties of nonlinear modes and  the specific forms of the confining
potential $V({\bf x})$ has not been discussed, previously.  This is a problem of significant practical importance since  high order modes
can be used for generation of nonlinear coherent structures, such as
for example solitonic trains in quasi-one-dimensional limit (see
e.g.~\cite{KevrekDyn}).

In this paper we study how the shape of the potential $V(x)$ governs the stability properties of
nonlinear modes. In our study and to simplify the analysis we will consider a
quasi-one dimensional geometry modeled by a one-dimensional GPE \cite{cigar-shape}
\begin{equation}
i\Psi_t=-\Psi_{xx}+V(x)\Psi-\sigma \Psi |\Psi|^2 .\label{GPEq_1}
\end{equation}
In that situation,  Eq. (\ref{StatGPEq}) becomes
\begin{equation}
\psi_{xx}+(\omega-V(x))\psi+\sigma \psi^3=0.
\label{MainEq}
\end{equation}
We will show that the  harmonic potential $V(x)=x^2$ corresponds to a
very peculiar situation related to the fact that in the linear limit
this potential  has an equidistant spectrum. We will discuss how
switching from the harmonic potential to an anharmonic one makes higher
nonlinear modes ``more stable'' and even a ``weak'' anharmonicity (say,
$V(x)=x^2+\kappa x^4$, $|\kappa|\ll 1$) is enough to change drastically
the stability properties of high-order nonlinear modes.

It is relevant to point out that the nonlinearity  introduces not
only mathematical difficulties for the study of the eigenmode
problem, but significantly diversifies the list of physically
relevant limiting cases. Indeed, the linear case has {\em two}
characteristic scales: the de Broglie wavelength and  the scale of
the potential. In our notations, the first one is $\lambdabar \sim
\max_x|\psi_x|/\max_x|\psi|$; the scale of the potential, roughly
speaking coincides with the classically allowed domain (in what
follows denoted as $x_d$). The nonlinearity introduces a {\em
third} relevant scale $\ell\sim 1/\max_x|\psi|$. For repulsive
nonlinearities this scale corresponds to the healing
length~\cite{Pitaev}, while in the case of attractive
nonlinearities it measures the width of a matter-wave soliton.
Therefore, the diversity of limiting cases of the nonlinear
eigenvalue problem is characterized by the interplay between the
parameters $\lambdabar$, $x_d$, and $\ell$.

The paper is organized as follows. In the  introductory
section~\ref{General} we describe the typical structure of the families
of nonlinear modes and formulate the stability problem. The results for
the harmonic potential are summarized  in Sec.~\ref{Harmonic}. Next, we
turn to the consideration of the anharmonic potentials $V(x)=x^2+\kappa
x^4$ (Sec.~\ref{UnHarmonic}) and $V(x)=x^4$ (Sec.~\ref{x4}) .
The last section~\ref{Conclusion} contains some additional
discussions and a summary of our results.

\section{Definitions and previous results}
\label{General}

\subsection{Branches of nonlinear modes}
\label{Branches}

 In the small amplitude limit $\psi\to 0$, the cubic term
in Eq. (\ref{MainEq}) can be neglected and the solutions can be approximated by the eigenfunctions
$\tilde\psi_n(x)$, $n=0,1,\ldots$ of linear eigenvalue problem
\begin{equation}
\psi_{xx}+(\omega-V(x))\psi=0. \label{LinMainEq}
\end{equation}
It is assumed that the potential  $V(x)$ is nonsingular, bounded from
below, and $V(x)\to\infty$ as $|x|\to\infty$, so that the spectrum of
(\ref{LinMainEq}) is discrete. Throughout this paper we will deal with
even potentials, $V(x)=V(-x)$. The real eigenfunctions of
(\ref{LinMainEq}), we denote them $\tilde\psi_n(x)$ with
$n=0,1,\ldots$,  constitute an orthonormal set
\begin{eqnarray}
\label{norma}
\langle \tilde\psi_n,\tilde\psi_m\rangle \equiv\int_{-\infty}^\infty\tilde\psi_n(x)\tilde\psi_m(x)~dx=\delta_{m,n}
\end{eqnarray}
(here $\delta_{m,n}$ is the  Kronecker delta).

It will be convenient to describe the families of nonlinear modes in terms
of bifurcation diagrams in the plane $(\omega,N)$ where
\begin{eqnarray}
\label{N}
N=\int_{-\infty}^\infty \psi^2(x)\,dx,
\end{eqnarray}
corresponds to the number of particles. Then, the respective
eigenvalues of (\ref{LinMainEq}), $\omega=\tilde{\omega}_n$,
$n=0,1,\ldots$, are the points of bifurcation where families of
nonlinear modes of Eq. (\ref{MainEq}), to be denoted as
$\Gamma_0,\Gamma_1,\Gamma_2,\ldots$, branch off from the zero
solution $\psi(x)\equiv 0$. The branching off takes place for both
cases, $\sigma=1$ (attractive nonlinearity) and $\sigma=-1$
(repulsive nonlinearity). Therefore we will label the branches of
nonlinear modes in the corresponding cases by $\Gamma_n^{(a)}$ and
$\Gamma_n^{(r)}$. At the same time, in the  statements which are
applicable to both cases $\sigma=\pm 1$, we omit the superscript
writing simply $\Gamma_n$. Two examples of diagrams for
$V(x)=x^2$ are shown in Fig. \ref{HarmFig}. Following
\cite{AgPres2002} we call the modes described above {\it modes
with linear counterpart}, since they can be viewed as modes of a
linear oscillator ``deformed'' by the action of the nonlinearity.

A simple analysis shows that in the vicinity of a bifurcation point, say
$\omega=\tilde{\omega}_n$, the small-amplitude solution of
Eq. (\ref{MainEq}) for the branch
$\Gamma_n$
can be described by the asymptotic expansions
\begin{subequations}
\label{AsEx}
\begin{eqnarray}
\psi_n(x) & =&\varepsilon\tilde{\psi}_n(x)+o(\varepsilon), \\
\omega_n&= & \tilde\omega_n-\varepsilon^2\sigma\Omega_n+o(\varepsilon^2),
\end{eqnarray}
\end{subequations}
where $\varepsilon\ll 1$ is a small parameter and the coefficient
$\Omega_n$ is given by
\begin{equation}
\Omega_n=\int_{-\infty}^\infty\tilde{\psi}_n^4(x)\ dx\,.
\label{omega2}
\end{equation}
From the physical point of view, $\Omega_n$ describes the two-body
interactions, and thus defines the characteristic scale $\ell$. Hence
the limit described by Eq. (\ref{AsEx}) in  physical terms can be
defined as $\ell\ll \lambdabar$. We observe that this limit can be
achieved not only due to small numbers of condensed particles,
expressed by the condition $\varepsilon\ll 1$ [due to the normalization
(\ref{norma})], but also for large enough $n$, since $\Omega_n\to 0$ as
$n$ grows. For instance, for the case $V(x)=x^2$ we have ~\cite{Szego}
\begin{eqnarray}
    \Omega_n={\cal O}(n^{-1/12}).
\end{eqnarray}

\subsection{Physical units}
\label{subsec:phys_unit}

Throughout this paper the nonlinear modes will be  characterized mainly in terms of the number of particles $N$ and the mode frequencies $\omega$. Since we will be interested in applications of our results to the BEC mean field theory while our analysis will be carried out in dimensionless units, before going into details we outline the link of our variables with experimentally feasible parameters. To this end we notice that the dimensionless form of Eq. (\ref{MainEq}) corresponds to the situation where the distance and time are measured respectively in units of $a_0$ and $2/\omega_0$, where $a_0$ is the longitudinal length of the trap and $\omega_0=\hbar/(m a_0^2)$, with $m$ being the atomic mass, is the "effective" longitudinal trap frequency (in the case of a parabolic potential it is the real frequency of  trap, while $a_0$ is the linear oscillator length). In the chosen scaling the energy is measured in the units $\hbar\omega_0/2$. Then it is a straightforward algebra to ensure that the link between the real, i.e physical, number of particles ${\cal N}$ and the norm of the solution $N$ (also referred to as number of particles) introduced by (\ref{N}) is given by the formula ${\cal N}=(a_\bot/4\sqrt{2}\pi |a_s|)N$, where $a_\bot$ is the transverse linear oscillator lengths and $a_s$ is the s-wave scattering length. In this last formula, as well as in the reduction of the 3D Gross-Pitaevskii equation to the 1D model (\ref{MainEq}), we have assumed that the trap is cigar-shaped, i.e. $a_\bot\ll a_0$, and that in the transverse direction the trap is harmonic. Now it is not difficult to estimate, that, for example for a trap with characteristic values $a_0\sim 1\,$mm, $a_\bot\sim 10\,\mu$m, and $a_s\sim 5\,$nm, a unit of the norm $N$ corresponds to ${\cal N}\sim 10^2$ atoms, or to the mean atomic density $\sim 0.4\cdot 10^9\,$cm$^{-3}$.

\subsection{Quasiclassical quantization}
\label{BSQC}

The limit of large $N$ and
 small local densities $|\psi|^2\ll 1$, denoted nonlinear WKB
approximation~\cite{KonKev}, allows for an analytical construction of
explicit solutions. Let us consider $V(x)= x^{2d}$, where $d$ is a
positive integer
 and assume that $N\gg 1$, or more precisely, that
$\delta=N^{-1-1/d}\ll 1$.  Bound states corresponding to
sufficiently large $\omega$, specifically $E=\omega N^{-2}\gtrsim
1$, correspond to an atomic cloud distributed over a large spatial
domain roughly determined by the classical turning points,
 $\pm x_d=\pm \omega^{1/2d}$.
Since $x_d$ grows with $\omega$, one can reach the levels
corresponding to sufficiently low densities of particles, i.e. the quasi-linear limit.

Returning to the definition of modes with linear counterpart,    the
arguments presented in this section allow one to conjecture, that {\em
only modes with a linear counterpart can exist in the limit $\omega\to
\infty$ at $N$ fixed}.

To describe that situation, we focus on the
repulsive case $\sigma<0$, introduce a new independent variable
$\zeta=\delta^{1/(1+d)}x$ and a renormalized macroscopic wave function
$\phi(\zeta)=\delta^{(d-1)/2(d+1)}\psi(x)$, and  rewrite Eq.
(\ref{MainEq}) as follows
\begin{eqnarray}
\label{WKB1} \delta^2\phi_{\zeta\zeta}+(E-\zeta^{2d})\phi-\delta
\phi^3=0.
\end{eqnarray}

Since now $\delta\ll 1$  Eq. (\ref{WKB1})  is a convenient
representation of the stationary eigenvalue problem for the
application of the nonlinear WKB approximation. Skipping details,
which can be found in \cite{KonKev}, here we present the equation
implicitly defining the diagram in the plane $(E,\delta)$:
\begin{widetext}
\begin{eqnarray}
\label{E} \displaystyle{
E^{\frac{1+d}{2d}}\approx\frac{\delta}{A_d}\left\{\pi\left(n+\frac
12\right)+ \frac{2B_d}{\sqrt{E} }\ln\left(\frac{32 E}{\delta}
\right) - C_d E^{-\frac{1+d}{2d}}+ \arg\left[\Gamma^2\left(1-
i\frac{B_d}{\sqrt{E}} \right)\right]\right\}. }
\end{eqnarray}
Here $n$ stands for the energy level and we introduced the constants
\begin{eqnarray*}
&& A_d=\left[\int_{-1}^{1}\sqrt{1-y^{2d}} dy\right]^{-1},\qquad B_d=
\frac{1}{2d} \left[\int_{-1}^{1}\frac{dy}{\sqrt{1-y^{2d}}}\right]^{-1},
\qquad \mbox{for $d=1,2,...$}
\\
&& C_d=  3B_d\sum_{k=1}^{d-1}\left\{\frac
12\cos\left(\frac{k\pi}{d}\right)\ln
\left[\frac{1-\cos\left(\frac{k\pi}{d}\right)}
{1+\cos\left(\frac{k\pi}{d}\right)}\right]-
\sin\left(\frac{k\pi}{d}\right)\arctan\left[\sin\left(\frac{k\pi}{d}\right)\right]\right\}\qquad
\mbox{for $d=2,3,...$}
\end{eqnarray*}
\end{widetext}
and $C_1=0$. $\Gamma(\cdot)$ is the standard notation for the
gamma function~\cite{AS}.

Solutions of the transcendental equation (\ref{E})  with respect
to the energy $E$ at fixed $\delta$ and $n$ give the eigenvalues
(energy levels). According to the previous discussion, when
$E\to\infty$ one recovers the WKB formula for the energy levels of
the potential $V(x)=x^{2d}$. While for the case of the harmonic
oscillator the  form of (\ref{E}) can be found in
Ref.~\cite{KonKev}, for the situation of our particular interest
below, $d=2$, the limit $E\to\infty$ (and thus $n\to\infty$) of
the nonlinear WKB equation (\ref{E}) acquires the form
\begin{eqnarray}
\label{WKB}
E_n^{3/4}&=&\frac{2\sqrt{2}}{3}K\left(\frac{1}{\sqrt{2}}\right)\delta\left[\pi\left(n+\frac 12 \right)
\right.
\nonumber \\
&+&
\frac{\ln n}{2\sqrt{2\pi}K\left(\frac{1}{\sqrt{2}}\right)\sqrt{n}}
+
\left.
{\cal O}\left(\frac{1}{\sqrt{n}}\right)\right],
\end{eqnarray}
where $K(\cdot)$ is the complete elliptic integral of the first kind~\cite{AS}.

Considering the last two terms in the expansion of the energy levels (\ref{WKB}) as a perturbation for $n$ large enough, and neglecting them in the leading order, in remaining part of the expression for $E_n$ one readily recognizes the familiar Bohr-Sommerfeld quantization condition. Thus formula (\ref{WKB}) can be viewed as the nonlinear generalization of the standard quasi-classical quantization well known in the quantum mechanics.

\subsection{The stability problem (general)}
\label{Stabil_Gen}

Consider now the stability problem for the modes corresponding to a
fixed branch $\Gamma_n$. Let $\psi_n(x)$ be a solution  of
Eq. (\ref{MainEq}) corresponding to $\omega_n$. Following the standard
procedure we represent $\Psi(x,t)=(\psi_n(x)+\xi(x,t))e^{-i\omega_n
t}$,  linearize the dynamical equation with respect to $\xi(x,t)$, and
arrive at the equation
\begin{eqnarray*}
i\xi_t=-\xi_{xx}-(\omega_n-V(x))\xi-\sigma\psi_n^2\left(2\xi+\xi^*\right),
\end{eqnarray*}
where the asterisk stands for the complex conjugation. Decomposing $\xi(x,t)$ into real
and imaginary parts, $\xi(x,t)=\chi(x,t)+i\varphi(x,t)$ we obtain
\begin{eqnarray}
\chi_t=-L_n^-~\varphi, \quad \varphi_t=L_n^+~\chi\label{chi_phi},
\end{eqnarray}
where
\begin{eqnarray*}
L_n^+=L_n+3\sigma \psi^2_n(x),\label{L+}
\quad
L_n^-=L_n+\sigma \psi^2_n(x),\label{L-}
\end{eqnarray*}
and
\begin{equation}
L_n=\frac{d^2\,}{dx^2}+\omega_n-V(x). \nonumber
\end{equation}
It follows from (\ref{chi_phi}) that  stability of the nonlinear mode is determined by
the spectrum of the following eigenvalue problem
\begin{equation}
L_n^-L_n^+~\zeta=\Lambda~\zeta.\label{L+L-}
\end{equation}
Since the operator $L^-_n$ is degenerate (the kernel of this operator
contains at least the function $\psi_n(x)$), the eigenvalue problem
(\ref{L+L-}) has  a zero eigenvalue. Then if the remainder of the spectrum of
$L_n^-L_n^+$ is {\it real} and {\it nonnegative} then the nonlinear
mode $\Psi(x,t)=e^{-i\omega_n t}\psi_n(x)$ is said to pass the linear stability
test. The presence of {\it negative} or {\it complex} eigenvalues in
the spectrum of $L_n^-L_n^+$ implies the linear instability of this
mode.

\subsection{The stability problem (small amplitude modes)}
\label{Stabil_small}

If $\omega$ lies close to a bifurcation point $\omega_n$ then the
spectrum of the operator $L_n^-L_n^+$ can be analyzed by means of
 the asymptotic expansions (\ref{AsEx}). Specifically,
\begin{eqnarray}
&&L_n^+={\cal
L}_n+\sigma\varepsilon^2(3\tilde{\psi}^2_n(x)-\Omega_n)+o(\varepsilon^2),\label{L_n^+}\\[2mm]
&&L_n^-={\cal L}_n+\sigma\varepsilon^2(\tilde{\psi}^2_n(x)-\Omega_n)+o(\varepsilon^2),\label{L_n^-}\\[2mm]
&& L_n^-L_n^+={\cal L}_n^2+\varepsilon^2\sigma
M_n+o(\varepsilon^2),\label{L+-}
\end{eqnarray}
where
\begin{eqnarray}
&&{\cal L}_n=\frac{d^2\,}{dx^2}+\tilde{\omega}_n-V(x),\label{cal_L}\\
&& M_n=3{\cal L}_n\tilde\psi^2_n(x)
+(\tilde\psi^2_n(x)-2\Omega_n){\cal L}_n.\label{M_n}
\end{eqnarray}
The operator ${\cal L}_n$ is self-adjoint and its spectrum consists of
eigenvalues $\lambda_{n,k}=\tilde{\omega}_n-\tilde{\omega}_k$,
$k=0,1,2,\ldots$. In the limit $\varepsilon=0$ one has the operator
$L_n^-L_n^+={\cal L}_n^2$ and its spectrum becomes
$\Lambda_{n,k}=(\tilde{\omega}_n-\tilde{\omega}_k)^2$,
$k=0,1,2,\ldots$. Since all  $\Lambda_{n,k}$ are real and
nonnegative, then if there are no multiple eigenvalues in this
spectrum, the small amplitude nonlinear modes are {\it linearly stable}
for both, repulsive and attractive nonlinearities. However, if the
spectrum of ${\cal L}_n^2$ includes multiple eigenvalues the stability
analysis implies the study of splitting of these eigenvalues when
passing from $\varepsilon=0$ to $0<\varepsilon\ll1$ (see
e.g.~\cite{Gelfand,Kato}).

\subsection{Krein signature}
\label{Krein}

Let $n$ be fixed and a pair $(\Lambda,\zeta(x))$ be a solution of
eigenvalue problem (\ref{L+L-}) where $\Lambda>0$ is a semi-simple
eigenvalue of $L_n^-L_n^+$ and the corresponding eigenfunction
$\zeta(x)$ is real. It is useful to assign to any such a pair
$(\Lambda,\zeta(x))$ the value
\begin{eqnarray*}
K={\rm sign}~\langle L_n^+\zeta(x),\zeta(x)\rangle
\end{eqnarray*}
called the {\it Krein signature}~\cite{MacKay1987}. As the solution
$\psi_n(x)$ of Eq. (\ref{MainEq}) varies along the branch $\Gamma_n$
together with $\omega_n$, the eigenvalues of the operator $L_n^+L_n^-$
also vary, but the Krein signature of any pair $(\Lambda,\zeta(x))$ is
conserved while there is no collision between eigenvalues. When a
collision between a pair of real positive eigenvalues takes place, they
can become complex {\it only} in the case when their Krein signatures
are opposite; otherwise these eigenvalues pass through each other both
remaining real. So, as $\omega_n$ varies along the branch  $\Gamma_n$
the interactions of eigenvalues with opposite Krein signatures may
affect the stability of modes in this branch.

The inverse statement  is also valid but {\it in a generic
situation} only~\cite{MacKay1987}: if the Krein signatures of
colliding eigenvalues are opposite then {\it generically} after
collision they become complex.
However additional symmetries
of the solution can destroy this picture: it will be shown that in some
cases eigenvalues with opposite Krein signatures can also  pass through
each other without causing instabilities.

\section{Results for the
harmonic potential $V(x)=x^2$.}
\label{Harmonic}

\subsection{General comments}

\begin{figure}
\epsfig{file=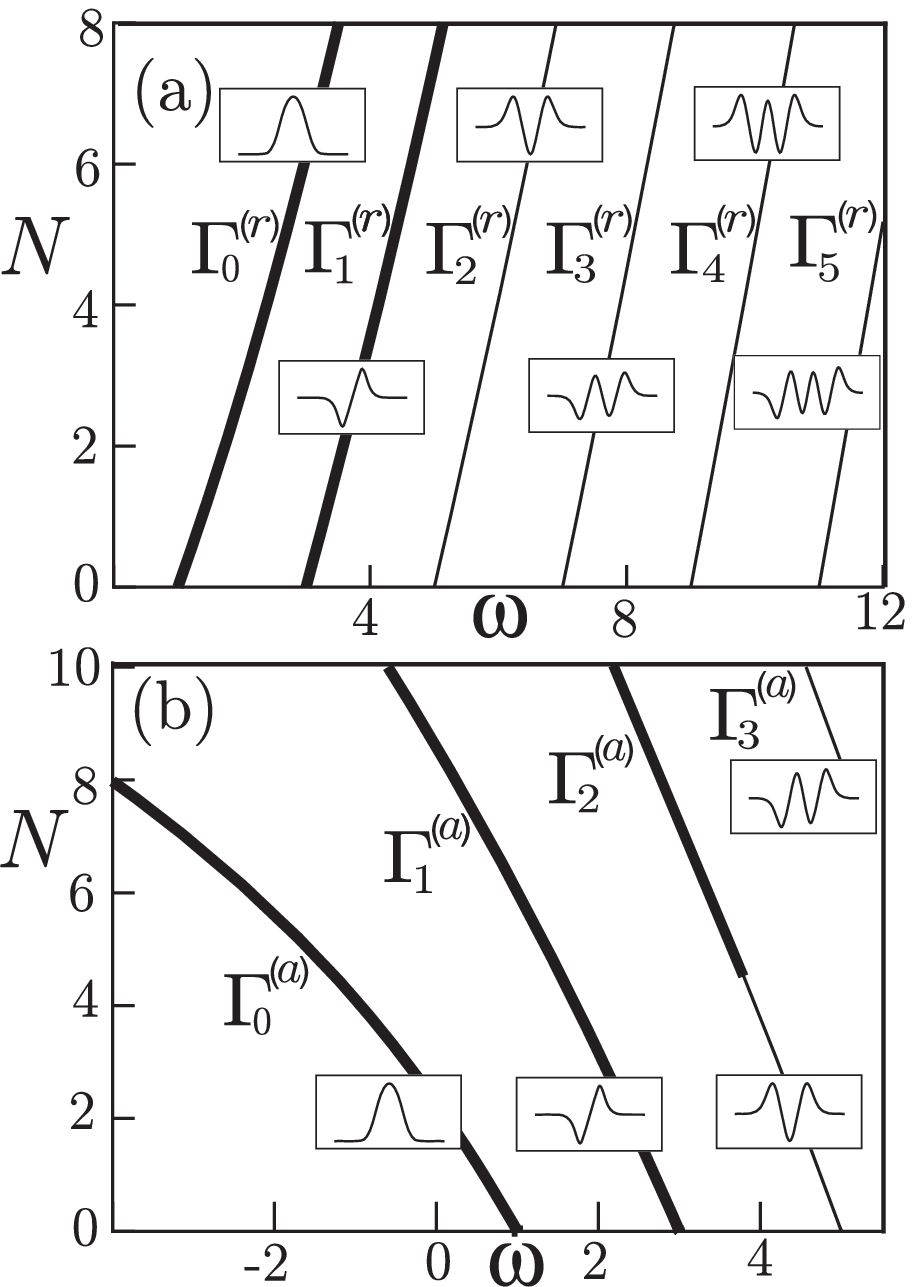,width=4.2cm}
\epsfig{file=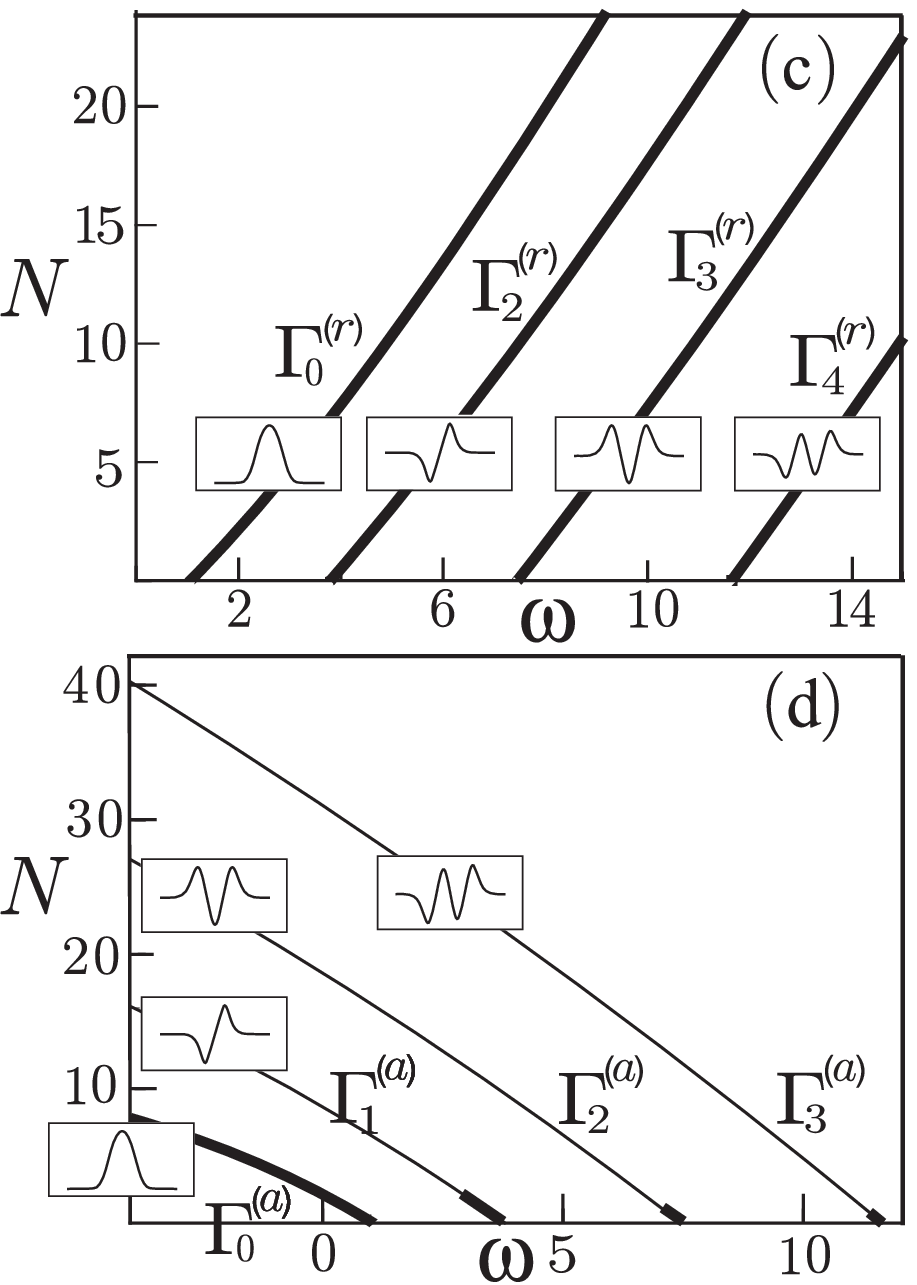,width=4.2cm}
\caption{ (a,b) The lowest branches of the nonlinear modes
of Eq. (\ref{MainEq}) for the potential $V(x)=x^2$ for repulsive
($\sigma=-1$) (a) and attractive ($\sigma=1$) (b) nonlinearities
respectively. (c,d) The same branches but for
$V(x)=x^4$. All the modes bifurcate from the  linear harmonic oscillator modes
 corresponding to the limit $N\to 0$. The fragments of curves corresponding to
stable solutions are shown in bold. } \label{HarmFig}
\end{figure}

In the case of the harmonic oscillator, where $V(x)=x^2$, the branches $\Gamma_n$, $n=0,1,2,\ldots$ are
depicted in Fig.~\ref{HarmFig} (a), (b).
It follows from Fig. \ref{HarmFig} that the branches $\Gamma_n$ are
represented by monotonic (at least for moderate values of $N$, $\omega$
and $n$) functions  $N(\omega)$. Previous numerical results \cite{AlfZez2007}
allow to conjecture that there are no solutions without linear
counterpart for this potential. It is known that the solutions
corresponding to the branches $\Gamma_0$ and $\Gamma_1$ in both,
attractive and repulsive cases, are stable (see, e.g. for instance
\cite{AlfZez2007,KevrekDyn} and \cite{PelKevr2007} for more detailed
analysis of perturbed solutions from $\Gamma_1$). Numerical
calculations show that the modes from $\Gamma_2^{(r)}$ (the repulsive
case) are unstable. In the attractive case the  family
$\Gamma_2^{(a)}$ corresponds to unstable modes for  $\omega$ close to the
bifurcation point $\omega_2=5$. More specifically, in
Ref.~\cite{AlfZez2007} the instability has been observed for
$\omega^*<\omega<5$ where $\omega^*\approx 3.83$. It has also been
found that for $\omega<\omega^*$ the mode is stable.

\subsection{Small-amplitude modes}
\label{SmallAm}

Let us now analyze the stability of the branch
$\Gamma_n$, for both attractive and repulsive cases when $\omega_n$ lies
close to the bifurcation point $\tilde{\omega}_n$. In the bifurcation point (linear limit)
 the solutions of the eigenvalue problem (\ref{LinMainEq})
are the pairs $(\tilde{\omega}_n,\tilde\psi_n(x))$, $n=0,1,\ldots$
where
\begin{subequations}
\label{Harm_eig}
\begin{eqnarray}
\tilde\omega_n & =& 2n+1,\\ \tilde\psi_n(x)& =& \frac 1{\sqrt{2^n
n!\sqrt{\pi}}} H_n(x)e^{-\frac12 x^2}
\end{eqnarray}
\end{subequations}
and $H_n(x)$ is $n$-th Hermite polynomial (see e.g.~\cite{AS}).

The stability of small amplitude solutions can be studied using the
formulas (\ref{L_n^+})-(\ref{M_n}). Let us start with the case
$\varepsilon=0$. The spectrum of the operator ${\cal L}_n$ is
equidistant and consists of the eigenvalues $\lambda_{n,k}=2(n-k)$ and
the corresponding eigenfunctions are given by $\tilde\psi_k(x)$,
$k=0,1,\ldots$. All the eigenvalues are simple and there is one zero
eigenvalue. The eigenvalues of the operator ${\cal L}^2_n$ are
$\Lambda_{n,k}=\lambda_{n,k}^2=4(k-n)^2$ and they correspond to the
same eigenfunctions $\tilde\psi_k(x)$, $k=0,1,\ldots$. This means that
the spectrum of ${\cal L}^2_n$ includes $n$ {\it double eigenvalues}
$\Lambda_{n,k}=4(n-k)^2$, $k=0,1,\ldots (n-1)$, one {\it simple zero
eigenvalue} and {\it infinitely many simple positive eigenvalues}. The
mechanism of emerging of double eigenvalues becomes transparent from
Table \ref{Table1}.  Each of the double eigenvalues
$\Lambda_{n,k}$
has an invariant subspace spanned by two functions,
$\tilde{\psi}_k(x)$ and $\tilde{\psi}_{2n-k}(x)$.
\begin{table}
\begin{tabular}{lllllllll}
\hline
&& 0-th & 1-st & 2-nd & 3-rd & 4-th & 5-th &$\ldots$\\
Eigenf-n && $\tilde\psi_0(x)$ & $\tilde\psi_1(x)$ & $\tilde\psi_2(x)$ &
$\tilde\psi_3(x)$ & $\tilde\psi_4(x)$ & $\tilde\psi_5(x)$ &\ldots\\[2mm]
\hline
${\displaystyle {\cal_L}_{1}}$& & 2 & 0 & -2 & -4 & -6 & -8 &$\ldots$\\[2mm]
${\displaystyle {\cal_L}_1^2}$& & \fbox{4} & 0 & \fbox{4} & 16 & 36 & 64 &$\ldots$\\[2mm]
\hline
${\displaystyle{\cal_L}_{2}}$& & 4 & 2 & 0 & -2 & -4 & -6 &$\ldots$\\[2mm]
${\displaystyle{\cal_L}_2^2}$& & \fbox{16} & \fbox{4} & 0 & \fbox{4} & \fbox{16} & 36 & $\ldots$\\[2mm]
\hline\end{tabular} \caption{Eigenvalues of ${\cal_L}_{1,2}$ and
${\cal_L}^2_{1,2}$. The boxes are used to emphasize the doubles eigenvalues}
\label{Table1}
\end{table}

Following Ref.~\cite{Kato} we consider  the $2\times 2$ matrices
\begin{eqnarray*}
\tilde{M}_{n,k}=\left(
\begin{array}{cc}
\langle M_n\tilde\psi_k,\tilde\psi_k\rangle & \langle
M_n\tilde\psi_k,\tilde\psi_{2n-k}\rangle
\\[2mm]
\langle M_n\tilde\psi_{2n-k},\tilde\psi_k\rangle & \langle
M_n\tilde\psi_{2n-k},\tilde\psi_{2n-k}\rangle
\end{array}
\right).
\end{eqnarray*}
If the eigenvalues of $\tilde{M}_{n,k}$ are $\mu^{(1)}_{n,k}$ and
$\mu^{(2)}_{n,k}$, $\mu^{(1)}_{n,k}\ne \mu^{(2)}_{n,k}$, then for
$\varepsilon\ll1$ the double eigenvalue $\Lambda_{n,k}$ of ${\cal
L}_n^2$ splits into two simple eigenvalues of $L_n^-L_n^+$:
$
\Lambda_{n,k}^{(j)}=\Lambda_{n,k}+\varepsilon^2\sigma\mu^{(j)}_{n,k}+o(\varepsilon^2)
$  where $j=1,2$.
Therefore, if the eigenvalues of any of the matrices
$\tilde{M}_{n,k}$, $k=0,\ldots,n-1$ are complex, the instability
of the small-amplitude solution $\Psi(t,x)=e^{-i\omega_n
t}\psi_n(x)$ takes place. It is important that since both,
repulsive ($\sigma=-1$) and attractive ($\sigma=1$) cases are
described by the eigenvalues of the same matrices
$\tilde{M}_{n,k}$, the complex eigenvalues of $\tilde{M}_{n,k}$
for some $k$ means the instability of small-amplitude modes in
both, repulsive and attractive cases.

Simple, but tedious algebra gives the  expressions for the
elements of the matrices $\tilde{M}_{n,k}$:
\begin{widetext}
\begin{eqnarray*}
\langle M_n\tilde\psi_k,\tilde\psi_k\rangle &=& \frac{8(n-k)}{\pi
2^{(n+k)}n!k!}\int_{-\infty}^\infty H_n^2(x)H_k^2(x)e^{-2x^2}~dx -
\frac{4(n-k)}{\pi 2^{2n}(n!)^2}\int_{-\infty}^\infty
H_n^4(x)e^{-2x^2}~dx,\\
\langle M_n\tilde\psi_k,\tilde\psi_{2n-k}\rangle &=&-\langle M_n\tilde\psi_{2n-k},\tilde\psi_k\rangle =\frac{4(n-k)}{\pi
2^{2n}n!\sqrt{k!(2n-k)!}}\int_{-\infty}^\infty
H_n^2(x)H_{2n-k}(x)H_k(x)e^{-2x^2}~dx,\\
\langle M_n\tilde\psi_{2n-k},\tilde\psi_{2n-k}\rangle &=&-\frac{8(n-k)}{\pi
2^{(3n-k)}n!(2n-k)!}\int_{-\infty}^\infty
H_n^2(x)H_{2n-k}^2(x)e^{-2x^2}~dx + \frac{4(n-k)}{\pi
2^{2n}(n!)^2}\int_{-\infty}^\infty H_n^4(x)e^{-2x^2}~dx.
\end{eqnarray*}
\end{widetext}
Using  Maple we calculated the eigenvalues of the matrix
$\tilde{M}_{n,k}$. The results are collected in Table \ref{Table2}
were we observe the following facts:
\begin{table}
\begin{tabular}{cccccccc}
\phantom{111}& $n=1$ & $n=2$ & $n=3$ & $n=4$ & $n=5$ & $n=6$ \\
\hline%
$k=0$ &%
$\begin{array}{c} 0.199\\ 0 \end{array}$ &%
C &%
C &%
$\begin{array}{c} 0.133\\ 0.404 \end{array}$ &%
$\begin{array}{c} -0.005\\ 0.618 \end{array}$ &%
$\begin{array}{c} -0.135\\ 0.816 \end{array}$\\
\hline%
$k=1$ &%
-- &%
$\begin{array}{c} 0.125\\ 0 \end{array}$ &%
C &%
C &%
C &%
$\begin{array}{c} 0.058\\ 0.473 \end{array}$\\
\hline
$k=2$ &%
-- &%
-- &%
$\begin{array}{c} 0.089\\ 0 \end{array}$ &%
C &%
C &%
C \\
 \hline
$k=3$ &%
-- &%
-- &%
-- &%
$\begin{array}{c} 0.068\\ 0 \end{array}$ &%
C &%
C \\
\hline
$k=4$ &%
-- &%
-- &%
-- &%
-- &%
$\begin{array}{c} 0.055\\ 0 \end{array}$ &%
C \\
\hline
$k=5$ &%
-- &%
-- &%
-- &%
-- &%
-- &%
$\begin{array}{c} 0.045\\ 0 \end{array}$ \\
\hline
\end{tabular}
\caption{Double eigenvalues of the matrices $\tilde{M}_{n,k}$.
Each cell of the table for $n>k$ contains either letter ``C'', meaning that
the eigenvalues are complex
 or two real eigenvalues.}\label{Table2}
\end{table}

(i) In columns 2 to 6 there is at least one letter ``C''  which
means instability of small-amplitude modes belonging to the respective
branch $\Gamma_n$, for both, attractive and repulsive nonlinearities.
We conjecture that the instability of small-amplitude modes takes
place for all branches $\Gamma_n^{(a)}$ and $\Gamma_n^{(r)}$ with $n\geq 2$.

(ii) In the case $n=1$ the mode is stable. That confirms the
results of \cite{KevrekDyn} (see  Fig.~1 there). It is interesting
that in the limit $\varepsilon=0$ the algebraic and geometric
multiplicities of the eigenvalue $\Lambda=4$  are both equal to 2.
At the same time two real eigenvalues emerging from the double
eigenvalue $\Lambda=4$ for $\varepsilon\ll 1$ have opposite Krein
signatures which do not correspond to the generic case (see
\cite{MacKay1987}).

(iii) When $n=k+1$ the matrix $\tilde M_{n,k}$ has a zero eigenvalue. This
reflects the fact that
\begin{equation}
\zeta(x)=\frac{d\psi_n(x)}{dx}, \label{Lam=4}
\end{equation}
is an eigenfunction of the operator $L_n^-L_n^+$ corresponding to
$\Lambda=4$ for {\it any} mode $\psi_n(x)$ belonging to any branch
$\Gamma_n^{(a)}$ or $\Gamma_n^{(r)}$.

\begin{figure}
    \includegraphics[scale=0.45]{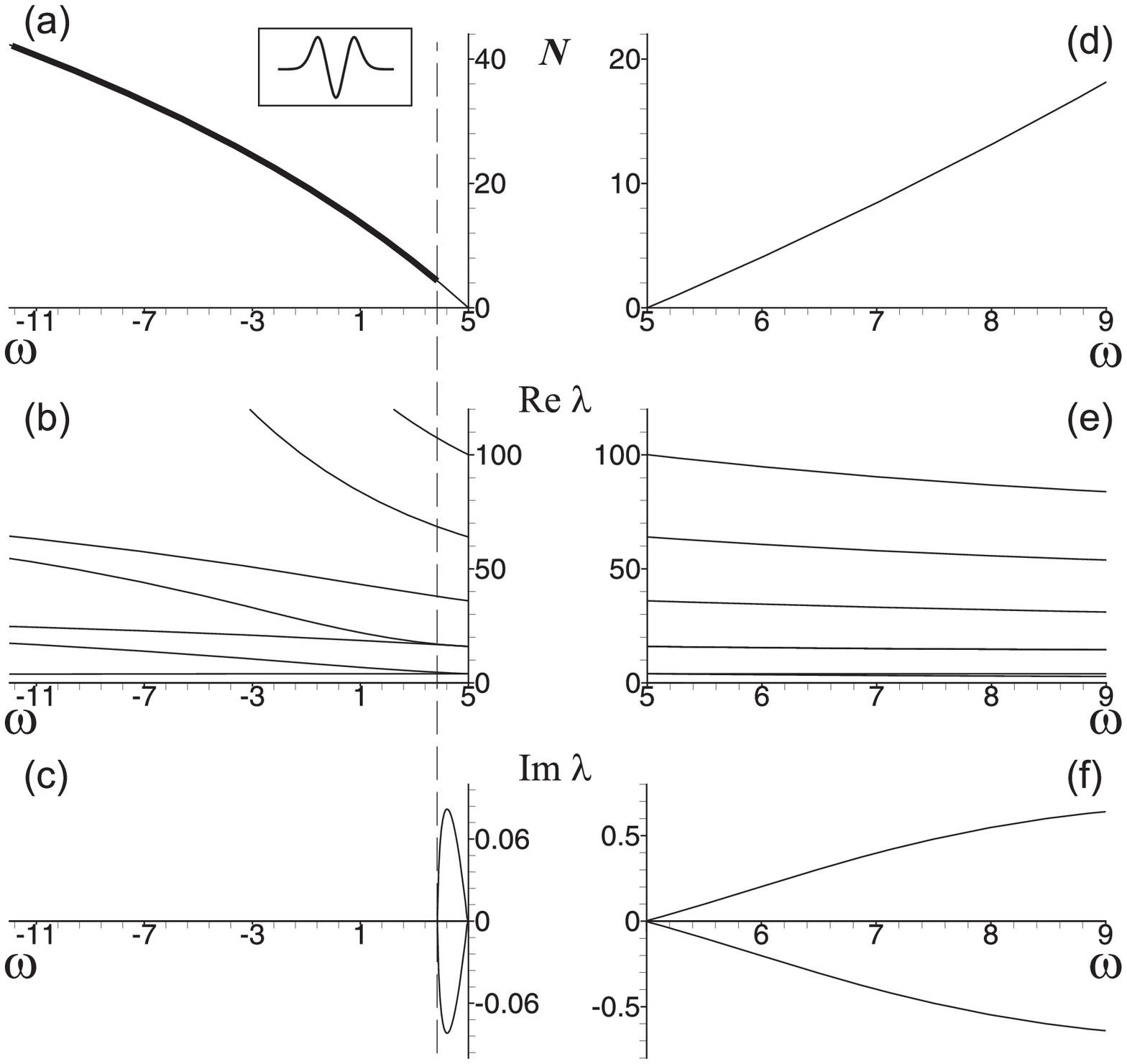}
    \caption{Branches $\Gamma_n^{(a,r)}$ and  plots of real and imaginary parts of eigenvalues
    $\Lambda$ of the operator $L_n^-L_n^+$ {\it vs}  $\omega$ for
    $V(x)=x^2$ and $n=2$.
    (a) $N$ {\it vs}
    $\omega$ for the branch $\Gamma_2^{(a)}$ (the part of the branch corresponding to
    stable solutions is shown in bold).
     (b) Real and  (c) imaginary parts of the eigenvalues
    $\Lambda$ for the attractive case. The splitting of the lowest eigenvalues $\Lambda=4$ and $\Lambda=16$
    of the linear problem is highlighted by the vertical dashed line. The nonzero imaginary part in panel (c)
    corresponds to the eigenvalues originated by $\Lambda=16$ of the linear problem for $\omega^*<\omega<5$
    (see the text).
    Plots (d)--(f) are analogous to  (a)--(c) but for the repulsive case. The branch studied is  $\Gamma_2^{(r)}$ and
    the eigenvalues originated by $\Lambda=16$ remain complex for the whole  interval of $\omega$ studied
    (see panel (f)), and therefore the mode is unstable.}
    \label{n=2x2}
\end{figure}

\begin{figure}
    \includegraphics[scale=0.45]{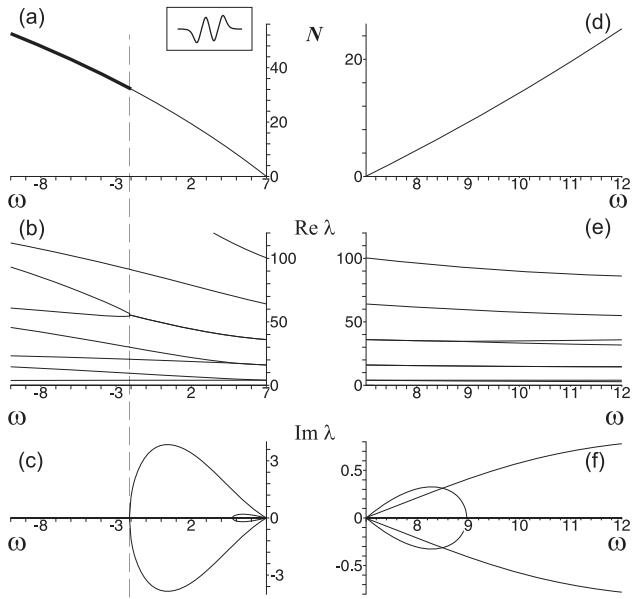}
    \caption{Branches $\Gamma_n^{(a,r)}$ and plots of the real and imaginary parts of the eigenvalues
    $\Lambda$
    {\it vs} $\omega$ for the potential  $V(x)=x^2$ and $n=3$.
    All the plots are organized in the same
    way as in  Fig.~\ref{n=2x2}.}
    \label{n=3x2}
\end{figure}

\subsection{Nonlinear modes of arbitrary amplitudes}
\label{NonSmall}

In order to study the linear stability of the nonlinear modes of a
finite amplitude we have first calculated the modes using a modified
shooting method developed in Ref. \cite{AlfZez2007} and used in Refs.
\cite{AlfZez2007,AZKV} to compute different families of nonlinear
modes. Then we have found numerically the eigenvalues of the operator
$L_n^-L_n^+$. To do so we have approximated the solution $\psi_n(x)$ on
a  grid, replacing the second derivatives in $L_n^{\pm}$ by
second-order finite differences and calculated the eigenvalues of the
resulting sparse matrix. The result is shown in Figs. \ref{n=2x2} and
\ref{n=3x2}. One can see that the families $\Gamma_2^{(a,r)}$  at
$\omega=5$ possess two double eigenvalues (Fig. \ref{n=2x2}, panels
b,c,e,f), $\Lambda=4$ (merged eigenvalues
$\Lambda_{2,1}=\Lambda_{2,3}$) and $\Lambda=16$, (merged eigenvalues
$\Lambda_{2,0}=\Lambda_{2,4}$). These eigenvalues split; in the case of
$\Lambda=4$ the resulting eigenvalues remain real and positive, one of
them corresponding to the exact solution (\ref{Lam=4}). In the case of
$\Lambda=16$, if the nonlinearity is attractive, there exist two
complex eigenvalues on the interval $\omega^*<\omega<5$,
$\omega^*\approx3.83$.   At $\omega=\omega^*$ these eigenvalues merge
and the mode becomes stable. Since the number of particles $N$ of the
mode grows when moving along the branch $\Gamma_2$, one can say that
there is  a {\it threshold on number of particles} for the stability of
the mode in attractive case. If the nonlinearity is repulsive, the two
complex eigenvalues do not disappear through all the region of
parameter $\omega$ investigated; therefore the mode remains unstable.
In the case of the family $\Gamma_3$, (Fig. \ref{n=3x2}, (a)-(f)) at
$\omega=7$ there are three double eigenvalues $\Lambda=4$, $\Lambda=16$
and $\Lambda=36$ which split. Then the scenario is similar to the case
of the family $\Gamma_2$. The eigenvalues originated by $\Lambda=4$
remain real and one of them corresponds to the exact solution
(\ref{Lam=4}). In the attractive case the mode becomes stable after the
two pairs of eigenvalues merge i.e. for $\omega\lesssim-2.10$; at this
point the second pair of the eigenvalues (both originated from
$\Lambda=36$) merges. In the repulsive case the mode remains unstable
through all the region of the parameter $\omega$  studied.

Summarizing, our results support  that  {\em high-order modes of GPE
with harmonic potential and attractive interactions, are stable when
the number of particles exceeds a threshold value} (different for each
branch), what corroborates our analysis on the quasi-linear behavior of
upper modes made in the begining of Sec.~\ref{BSQC}. In the repulsive
case, high-order modes of GPE with a harmonic potential are unstable.
Our result contradicts those
 of \cite{Carr2001} where it was claimed that stable modes exist for both signs
of nonlinear term.

\begin{figure}
    \includegraphics[scale=0.45]{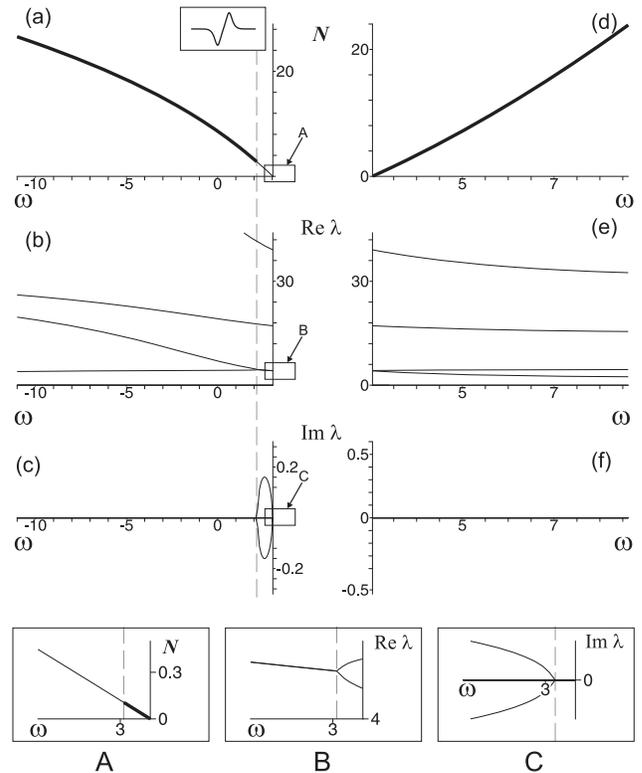}
    \caption{Branches $\Gamma_n^{(a,r)}$ and plots of the real and imaginary parts of eigenvalues
    $\Lambda$ of the operator $L_n^-L_n^+$ {\it vs} $\omega$ for  $V(x)=x^2+0.01x^4$ and
    $n=1$.
    Plots (a)--(f) are organized as in  Fig.~\ref{n=2x2}. In the attractive case the
    dashed line  on the plots (a)--(c)  marks the upper boundary of the
    instability window. The lower boundary of this instability window is
    very close to $N=0$, and not visible on the scale of the plots (a)--(f).
    Plots A, B and~C show
    the branch $\Gamma_1^{(a)}$ and the real and imaginary parts of eigenvalues
    $\Lambda_{1,0}$ and $\Lambda_{1,2}$ {\it vs} $\omega$ close to
    the linear limit  $\omega = \tilde{\omega}_1$ (i.e. $N=0$) with magnification.
    On the plots  A, B and C the lower boundary of the instability window is 
    marked with a~dashed line.}
    \label{n=1x2+}
\end{figure}
\begin{figure}
    \includegraphics[scale=0.45]{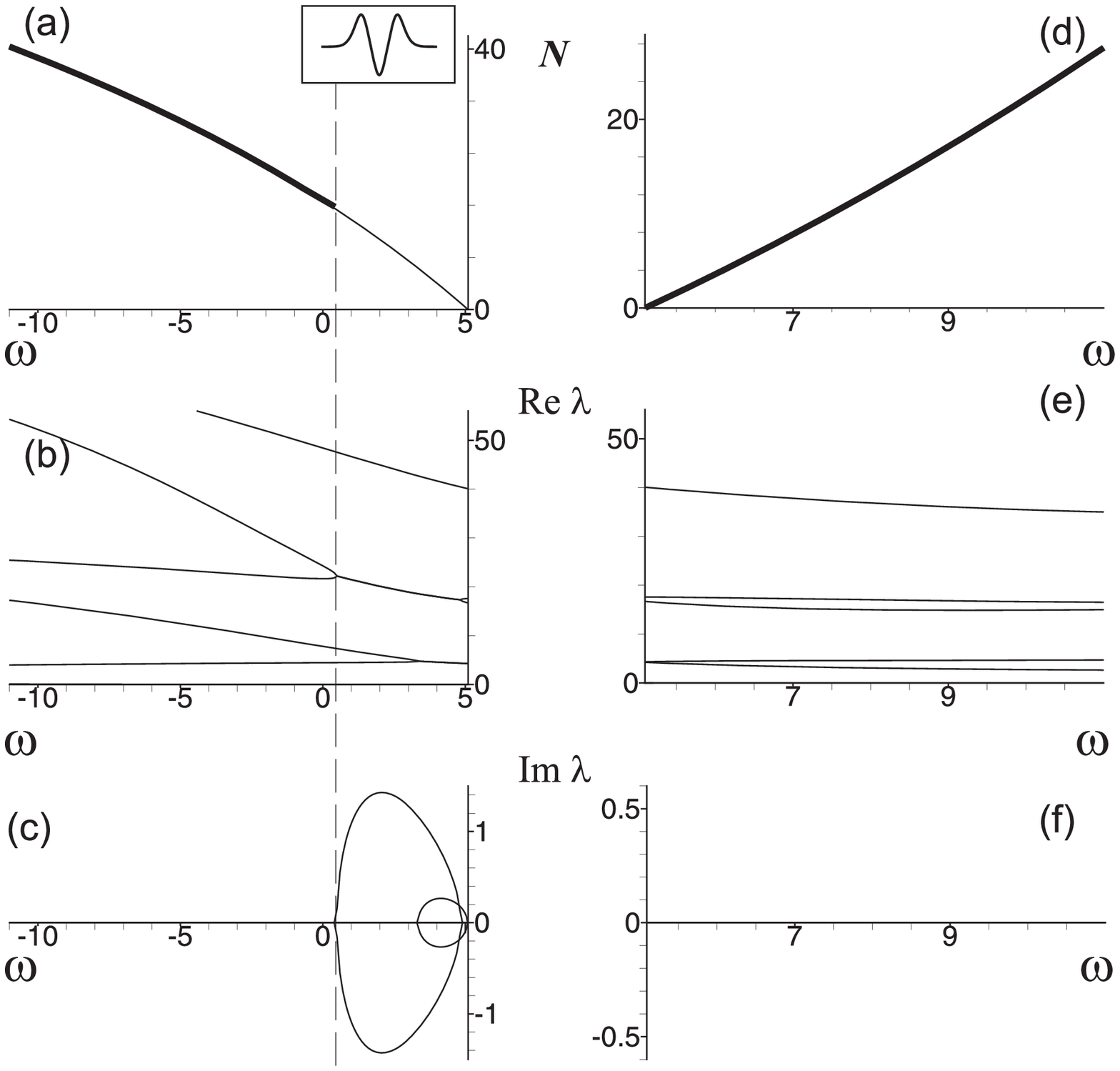}
    \caption{Branches $\Gamma_n^{(a,r)}$ and plots of the real and imaginary parts of eigenvalues
    $\Lambda$ of the operator $L_n^-L_n^+$ {\it vs} $\omega$ for
    $V(x)=x^2+0.01x^4$ and $n=2$.  In attractive case the lower boundary is
    very close to $N=0$, and not visible on the scale of the figure. All the plots are organized  as in Fig.~\ref{n=2x2}.}
    \label{n=2x2+}
\end{figure}
\begin{figure}
    \includegraphics[scale=0.45]{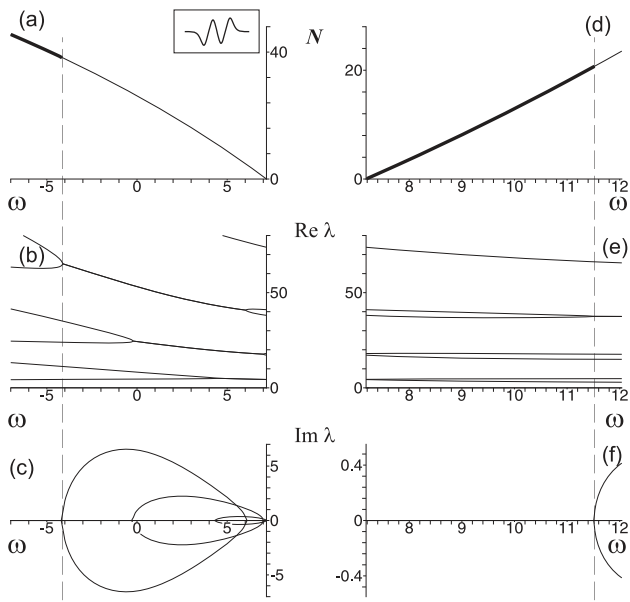}
    \caption{Branches $\Gamma_n^{(a,r)}$ and plots of the real and imaginary parts of eigenvalues
    $\Lambda$ of the operator $L_n^-L_n^+$ {\it vs} $\omega$ for
    $V(x)=x^2+0.01x^4$ and $n=3$. In attractive case the lower boundary is
    very close to $N=0$, and not visible on the scale of the figure.
    All the plots are organized as  in Fig.~\ref{n=2x2}.}
    \label{n=3x2+}
\end{figure}

\section{Anharmonic potentials (I) : Small perturbations of a harmonic potential}\label{UnHarmonic}
 \label{x2+Smallx4}

Now we turn our attention to the GPE with a harmonic potential
perturbed by a quartic term, $V(x)=x^2+\kappa x^4$, $0<|\kappa|\ll1$.
In this case the eigenvalues $\tilde{\omega}_n$ and eigenfunctions
$\tilde{\psi}_n(x)$, $n=0,1,\ldots$, for the linear problem
(\ref{LinMainEq}) can be found numerically or by means of asymptotic
procedures \cite{BendOrsz}. Also one have to employ numerics for the
construction of the branches of nonlinear modes $\Gamma_n^{(a)}$ and
$\Gamma_n^{(r)}$; which are similar to the corresponding branches for
harmonic potential case except for small deformations. As in the purely
harmonic case, the branches $\Gamma_n^{(a,r)}$ are monotonic (at least
for moderate values of $N$, $\omega$ and $n$) and can be parametrized
by any of the parameters $N$ and $\omega$. However, the spectrum
$\tilde{\omega}_n$ is no longer equidistant, which has several
important consequences as we will discuss in what follows.

\subsection{Small amplitude modes}
\label{SmallAm_x^4}

The stability of small amplitude modes belonging to the branches $\Gamma_n^{(a,r)}$
in the case of  the anharmonic potential can be also
studied by means of asymptotic expansions (\ref{L_n^+})-(\ref{M_n}).
Consider the case $\varepsilon=0$. The spectrum of the operator ${\cal
L}_n^2$ now consists of simple eigenvalues
 $\Lambda_{n,k} = (\tilde{\omega}_k- \tilde{\omega}_n)^2$,
$k=0,1,\ldots$, and the Krein signature of the eigenvalue $\Lambda_{n,k}$ is
$K_{n,k} = {\rm sign}(n-k)$. Therefore, if $\kappa$ is small the spectrum
of  ${\cal L}_n^2$ contains $n$ pairs of close eigenvalues of opposite
Krein signatures, $\Lambda_{n,k}$ and $\Lambda_{n,2n-k}$, $k=0,1,..,n-1$,
which are, nevertheless, different. The spectrum contains also a zero eigenvalue
and an infinite sequence of increasing positive eigenvalues. Therefore,
generally speaking, the spectrum $\Lambda_{n,k}$, $k=0,1,\ldots$ does not contain
multiple eigenvalues.

When passing from $\varepsilon=0$ to $0<\varepsilon\ll1$ the
eigenvalues $\Lambda_{n,k}$ of the operator $L_n^-L_n^+$ vary continuously.
Therefore, one can expect the stability of the mode until two
eigenvalues with opposite Krein signature merge. So, the small
amplitude modes {\it are expected to be stable} for all the branches
$\Gamma_n^{(a)}$ and $\Gamma_n^{(r)}$.

\subsection{Nonlinear modes of arbitrary amplitude: $\kappa>0$.}
\label{NonSmall_x4}

In order to study the linear stability of the nonlinear modes of a finite
amplitude we have first calculated the spectrum of the operator $L_n^-L_n^+$
numerically, concentrating on the branches $\Gamma_n^{(a,r)}$
for $n = 0,1,...,4$. The obtained general picture appears to be very different
from that of the harmonic potential.
The plots of real and imaginary parts of the eigenvalues $\Lambda$
{\it vs} $\omega$ for the branches $\Gamma_1$, $\Gamma_2$ and
$\Gamma_3$ in both attractive and repulsive cases are shown in
Figs.~\ref{n=1x2+}, \ref{n=2x2+}, and \ref{n=3x2+}.

The {\it general property} of  all the cases considered is that
the instability of the nonlinear modes
occurs {\it only}
due to collisions of  pairs of eigenvalues which are continuation
of the respective eigenvalues $\Lambda_{n,k}$ and $\Lambda_{n,2n-k}$ in the linear limit,
i.e. of the eigenvalues of the operator ${\cal L}_n^2$. In what follows we call them
{\it $(\Lambda_{n,k},\Lambda_{n,2n-k})$-pairs}. The eigenvalues in these
pairs have opposite Krein signatures.

Then, the other numerical results can be structured as follows.

\paragraph{ Repulsive nonlinearity.} The ground state modes corresponding to
the branch $\Gamma_0^{(r)}$ are stable, since there are no
$(\Lambda_{n,k},\Lambda_{n,2n-k})$-pair in the spectrum of
$L_0^-L_0^+$. The modes from the next branch, $\Gamma_1^{(r)}$, in the
limit of strong nonlinearity correspond to ``dark'' soliton modes;
there is one $(\Lambda_{n,k},\Lambda_{n,2n-k})$-pair in the spectrum of
$L_1^-L_1^+$, but no collisions of eigenvalues have been found when
tracing the modes of this branch within the range of parameter $N$
where it has been investigated (see Fig. \ref{n=1x2+}, right panels).
Therefore we conclude that the modes from $\Gamma_1^{(r)}$ are {\it
stable}. A similar situation takes place for the branch
$\Gamma_2^{(r)}$ where two $(\Lambda_{n,k},\Lambda_{n,2n-k})$-pairs in
the spectrum of $L_2^-L_2^+$ present: no collision has been observed
within the range of parameter $N$ under consideration (see
Fig.~\ref{n=2x2+}, right panels). However this is not the case for
higher branches. For instance, the collisions of eigenvalues has been
observed for nonlinear modes from $\Gamma_3^{(r)}$. The spectrum of the
operator $L_3^-L_3^+$ includes three
$(\Lambda_{n,k},\Lambda_{n,2n-k})$-pair and a collision of one of them
(highest) at some large enough value of $N$ has been seen (see
Fig.~\ref{n=2x2+}, right panels). After the point of collision (i.e.
for greater values of $N$) the pair of collided eigenvalues become
complex which means the instability of corresponding nonlinear modes. A
similar situation takes place for the branch $\Gamma_4^{(r)}$. In
general, this points out to the fact that the instability of higher
modes, generically, takes place, if the number of particles $N$ exceed
some threshold value, which is particular for each branch
$\Gamma_n^{(r)}$. The existence of the threshold value for the branches
$\Gamma_1^{(r)}$ and $\Gamma_2^{(r)}$ which we have not found in our
numerical investigation needs more delicate analysis.

\begin{figure}
    \includegraphics[scale=0.45]{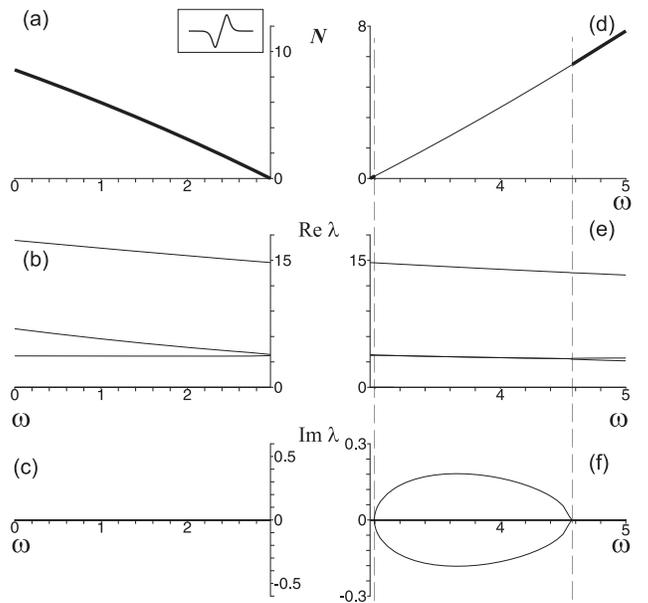}
    \caption{Branches $\Gamma_n^{(a,r)}$ and plots of the real and imaginary parts of eigenvalues
    $\Lambda$ of the operator $L_n^-L_n^+$ {\it vs} $\omega$ for
    $V(x)=x^2-0.01x^4$ and $n=1$. All the plots are organized  as in Fig.~\ref{n=2x2}.}
    \label{n=1x2-}
\end{figure}

\begin{figure}
    \includegraphics[scale=0.45]{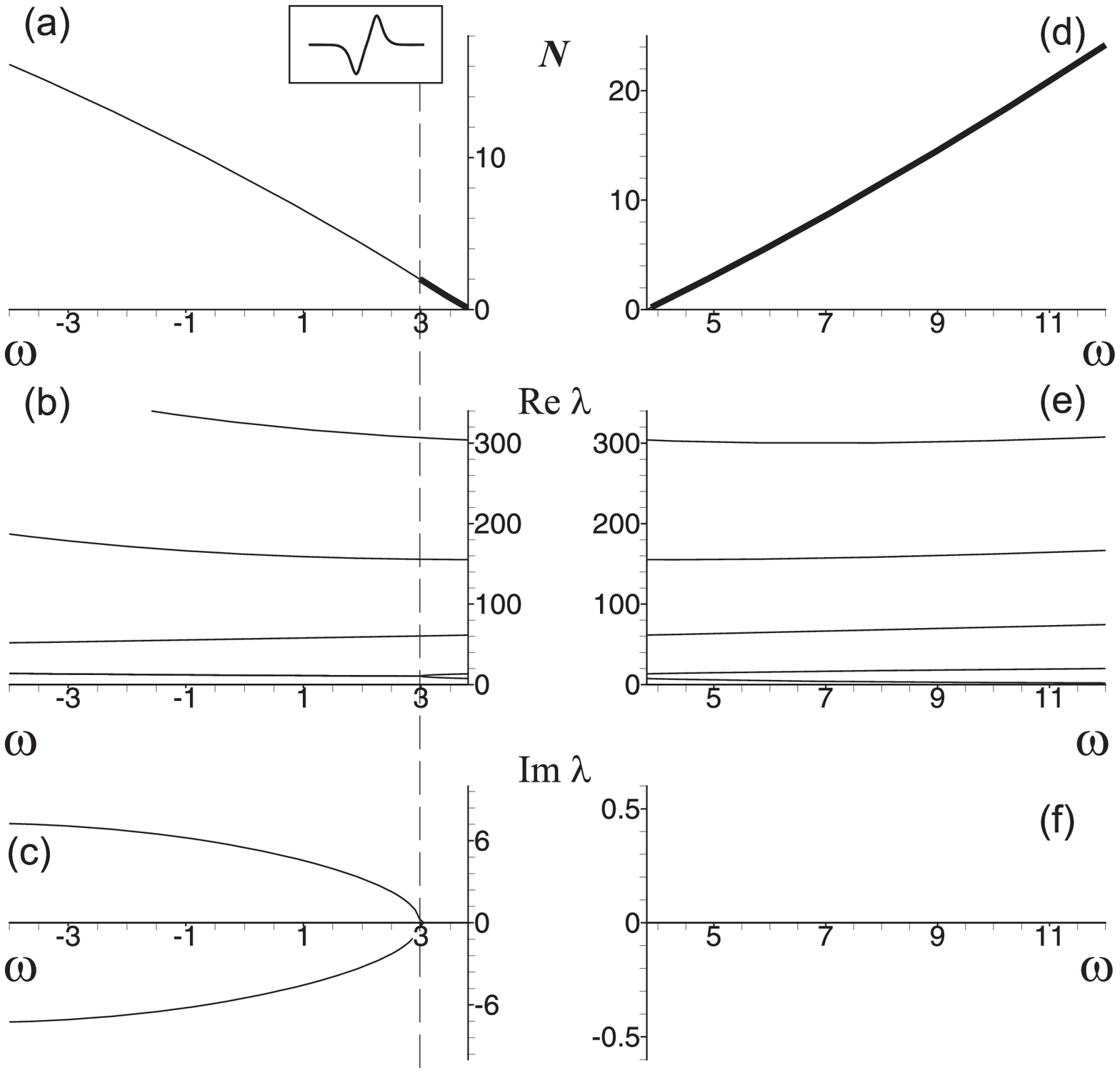}
    \caption{Branches $\Gamma_n^{(a,r)}$ and the plots of real and imaginary parts of eigenvalues
    $\Lambda$ of the operator $L_n^-L_n^+$ {\it vs} $\omega$ for
    $V(x)=x^4$ and $n=1$.
    All the plots are organized in the same
    manner as  in the Fig.~\ref{n=2x2}.}
    \label{n=1x4}
\end{figure}
\begin{figure}
    \includegraphics[scale=0.45]{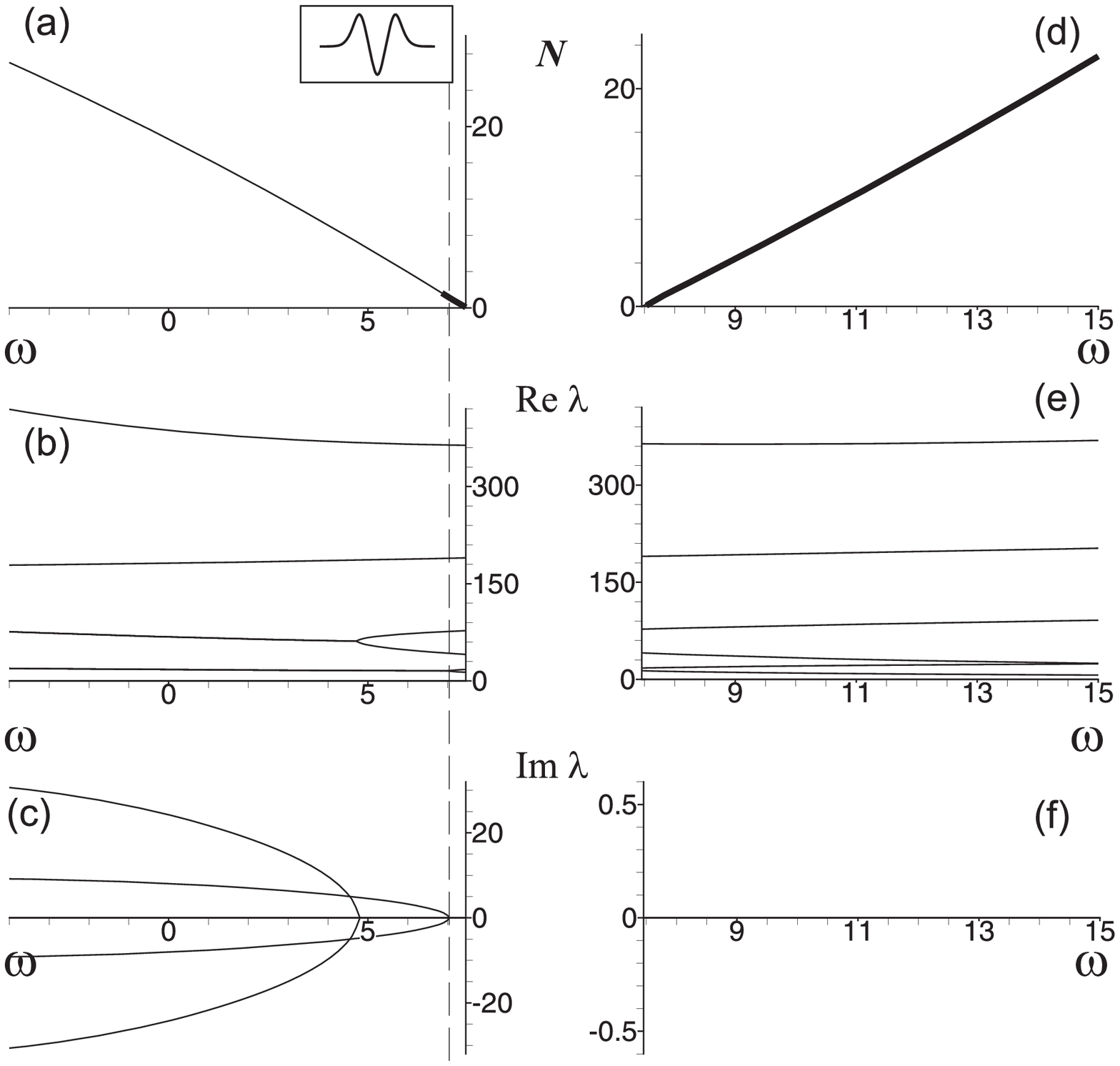}
    \caption{Branches $\Gamma_n^{(a,r)}$ and the plots of real and imaginary parts of eigenvalues
    $\Lambda$ of the operator $L_n^-L_n^+$ {\it vs} $\omega$ for  $V(x)=x^4$ and
    $n=2$.  
    All the plots are organized in the same
    manner as  in the Fig.~\ref{n=2x2}.}
    \label{n=2x4}
\end{figure}
\begin{figure}
    \includegraphics[scale=0.45]{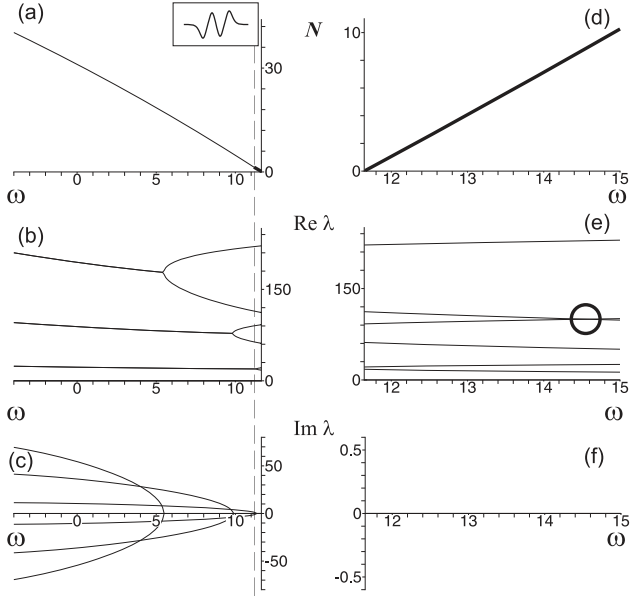}
    \caption{Branches $\Gamma_n^{(a,r)}$ and the plots of real and imaginary parts of eigenvalues
    $\Lambda$ of the operator $L_n^-L_n^+$ {\it vs} $\omega$ for  $V(x)=x^4$ and
    $n=3$.
    All the plots are organized in the same
    manner as  in the Fig.~\ref{n=2x2}. The bold circle on the plot~(e)
    highlights the collision of the eigenvalues of opposite Krein
    signatures which does not lead to instability.}
    \label{n=3x4}
\end{figure}

\paragraph{ Attractive nonlinearity.} The ground state modes
 of the branch $\Gamma_0^{(a)}$ are stable. For the branch
$\Gamma_1^{(a)}$, there is one
$(\Lambda_{n,k},\Lambda_{n,2n-k})$-pair in the spectrum of
$L_1^-L_1^+$ (see Fig. \ref{n=1x2+}, left panels). Then there
exist two bifurcation values of the parameter $N$ (the number of
particles). When $N$, increasing, reaches the first bifurcation
value, the eigenvalues of this pair collide and become complex.
For the values of $N$ below this threshold, the mode is stable;
however, for the potential under consideration the first
bifurcation value is very tiny, so it is not visible in Fig.
\ref{n=1x2+}. Then, when $N$ reaches the second bifurcation value,
the complex eigenvalue collide again and become real. So, we can
conclude that the modes of $\Gamma_1^{(a)}$ are {\it stable}
unless the number of particles $N$ belongs to an {\it instability
window}; the lower bound of this window is close to zero (but
separated from zero). The size of the window of instability
increases when $\kappa$ grows and both bounds of this window are
quite sensitive to variation of $\kappa$. A similar situation has
been observed for higher branches of nonlinear modes. For
instance, the spectrum of the operator spectrum of $L_2^-L_2^+$,
the branch $\Gamma_2^{(a)}$, includes two
$(\Lambda_{n,k},\Lambda_{n,2n-k})$-pairs. As $N$ grows both of
them undergo the same evolution: they become complex at first
bifurcation value of $N$ and return to be real at the second
bifurcation value (see Fig. \ref{n=2x2+}, left panels). The
interval with respect to $N$ between the first bifurcation value
for the pair $(\Lambda_{2,1},\Lambda_{2,3})$ and second
bifurcation value for the pair $(\Lambda_{2,0},\Lambda_{2,4})$
represents {\it the window of instability}. Upper boundary of this
instability window is marked by dashed line in Fig. \ref{n=2x2+}.
However, since the first bifurcation value for the pair
$(\Lambda_{2,1},\Lambda_{2,3})$ (lowest curve in Fig.
\ref{n=2x2+}, panel (b)) is very tiny, so the lower boundary of
instability window cannot be separate from zero in Fig.
\ref{n=2x2+}. Therefore, the modes from $\Gamma_2^{(a)}$ are
stable if $N$ does not belong to instability window. This
situation, probably, is generic for other higher branches.

To confirm our results on the stability of nonlinear high-order
modes we  also have performed a series of direct numerical
simulations of their evolution, perturbed by a random perturbation
of 5\% amplitude of the mode. Thus we have simulated the evolution
of initial data of the form $\psi_0(x) = \psi_n(x) \left[1+
r(x)\right]$ with $r(x)$ a white noise of maximum amplitude 0.05.
The subsequent dynamics of the modes under Eq. (\ref{GPEq_1}) was
computed using a second order in time split-step pseudospectral
scheme  discretized in space using trigonometric polynomials (via
the FFT). In all the cases studied, which included most of the
branches presented here our test verified the predictions based on
linear stability analysis.

\subsection{Nonlinear modes of arbitrary amplitude: $\kappa<0$.}
\label{NonSmall_x4-}

Let us briefly summarize the stability results for the potential
$V(x)=x^2+\kappa x^4$, $0<|\kappa|\ll 1$ and $\kappa<0$.  In this case
only a {\it finite} number of  branches $\Gamma_n^{(a,r)}$ can exist,
since there is a finite number of discrete eigenvalues for
Eq.(\ref{LinMainEq}).
These branches can be found numerically.  The linear stability
analysis performed for the potential $V(x) = x^2-0.01x^4$ shows
(see Fig.~\ref{n=1x2-}) that  all solutions of the branch
$\Gamma_1^{(a)}$ which we have considered are linearly stable. On
the other hand, the modes of $\Gamma_1^{(r)}$  are also stable,
except some instability window situated close to the point of
branching. This is in contrast with the case $\kappa>0$, since in that
case the instability window is situated on the branch
$\Gamma_1^{(a)}$ but not $\Gamma_1^{(r)}$. The solutions of the
branch $\Gamma_2^{(a)}$ are also linearly stable  whereas the
solutions of the branch $\Gamma_2^{(r)}$ are stable only in the
vicinity of the branching point, i.e. only for $N\ll 1$. For
the branches $\Gamma_3^{(a,r)}$ the picture of
stability/instability becomes more complex.

\section{Anharmonic potentials (II): Potential $V(x)=x^4$} \label{x4}

In order to show that the results for anharmonic potentials are
quite generic, let us consider GPE with  $V(x)=x^4$.
Again, the eigenvalues, $\tilde{\omega}_n$,  and eigenfunctions
$\tilde{\psi}_n(x)$, $n=0,1,\ldots$, for the linear problem
(\ref{LinMainEq}) cannot be obtained exactly. The branches of
nonlinear modes $\Gamma_n^{(a)}$ and $\Gamma_n^{(r)}$ which have
been found numerically for $n=0,1,2,3$ are similar to the
corresponding branches for harmonic potential case. These branches
are monotonic (at least for moderate values of $N$, $\omega$ and
$n$) and can be parametrized by any of the parameters $N$,
$\omega$.

Due to the reasons discussed above,  the small amplitude solutions of
GPE with the potential $V(x)=x^4$ generically {\it are stable} for
both, attractive and repulsive nonlinearities. In order to study the
stability of nonlinear modes in general, we have analyzed the
bifurcations of eigenvalues of $L_n^-L_n^+$ when $\psi_n(x)$ varies
along the families $\Gamma_n^{(a)}$ and $\Gamma_n^{(r)}$. Our numerical
investigation shows that, as in the case of weak anharmonicity,  the
instability of nonlinear modes from $\Gamma_n$  occurs {\it only} due
to collisions of those pairs of eigenvalues which are continuation of
eigenvalues $\Lambda_{n,k}$ and $\Lambda_{n,2n-k}$ in linear limit i.e.
in the spectrum of ${\cal L}_n^2$. These eigenvalues $\Lambda_{n,k}$
and $\Lambda_{n,2n-k}$ are of opposite Krein signature. Again, let us
refer to these pairs of eigenvalues as
$(\Lambda_{n,k},\Lambda_{n,2n-k})$-pairs.

Qualitatively, the stability picture for the case $V(x)=x^4$ is
the same as in the case $V(x)=x^2+\kappa x^4$, $\kappa>0$, except
for some minor differences.  The  first one is that we have not
found the threshold of instability in $N$ in the case of repulsive
nonlinearity for the modes in the branches $\Gamma_n^{(r)}$,
$n=0,1,2,3$. Therefore, these modes are stable in all the
parameter range studied. Second, we have not found an upper bound
of instability window in $N$ in the case of attractive
nonlinearity for the modes from $\Gamma_n^{(a)}$, $n=1,2,3$. We
believe that it is a technical problem related to the finiteness
of the region studied but we think an upper bound of this
instability window exists. The plots of real and imaginary parts
of eigenvalues $\Lambda$ of the operator $L_n^-L_n^+$ versus
$\omega$ for the branches $\Gamma_1$, $\Gamma_2$ and $\Gamma_3$
for attractive and repulsive nonlinearities
 are shown in Figs. \ref{n=1x4}, \ref{n=2x4}, and \ref{n=3x4}.

It is interesting to mention that we have also observed
collisions of eigenvalues with opposite Krein signatures which
belong to different $(\Lambda_{n,k},\Lambda_{n,2n-k})$-pairs. In
our case they did not lead to instability since they remained real
after the collision (see for example Fig. \ref{n=3x4}, panel (e)).
This phenomenon does not correspond to a generic situation; it is
caused by the opposite parity of colliding eigenfunctions (one of
them was odd, while the other one was even).

We have also performed a set of direct numerical simulations  of the
evolution of perturbed stationary modes in Eq. (\ref{GPEq_1}) as
described in Sec. \ref{NonSmall_x4}. The outcome of those simulations
confirms the results of the linear stability analysis.

\begin{widetext}

\begin{table}[h]
\begin{tabular}{||p{3.5cm}|p{2cm}|p{5.5cm}|p{5cm}||}
    \hline\hline
     & $V(x)=x^2$
     & $V(x)=x^2+\kappa x^4$, $\kappa>0$
     & $V(x)=x^4$
     \\
    \hline\hline
     0-th mode,
     (ground state),
     $\sigma=\pm 1$
     & {\bf stable} & {\bf stable} & {\bf stable}\\
    \hline\hline
     1-st mode, $\sigma=1$ small amplitude limit
     & {\bf stable}
     & {\bf stable}
     & {\bf stable}
    \\
    \hline
     1-st mode ("bright soliton"), $\sigma=1$  general case
     & {\bf stable}
     &  {\bf unstable}, if $N$ belongs to some
   ``instability window''.   The lower bound of this ``window'' is separated from zero.
   The size of the ``window'' grows
    with $\kappa$.  Otherwise {\bf stable}.
    &  {\bf unstable}, if   $N$ belongs to some
   ``instability window'', with lower bound separated from zero.
   Otherwise {\bf stable}.
    \\
   \hline\hline
     1-st mode,  $\sigma=-1$
    small amplitude limit  & {\bf stable} &
    {\bf stable} & {\bf stable}\\
    \hline
     1-st mode,(''dark soliton'') $\sigma=-1$,  general case
     & {\bf stable} &  {\bf stable} for all
   $N$ which have been considered.  &
    {\bf stable} for all $N$ which have been considered.
    \\
 \hline\hline
     Higher modes, $\sigma=1$  small amplitude limit  & {\bf unstable} & {\bf stable} &
   {\bf stable}\\
   \hline
   Higher modes, $\sigma=1$, general case
   & {\bf stable},  if $N$ exceeds a threshold.
   & {\bf unstable}, if  $N$ belongs to some ``instability window'', with lower bound separated from zero.
    The size of the ``window'' grows  with $\kappa$. Otherwise {\bf stable}.
    & {\em Hypothetically,} {\bf unstable}, if   $N$ lies in a large ``window'' of instability.
    The upper boundary of this ``window'' has not been found in our numerics.\\
    \hline\hline
    Higher modes, $\sigma=-1$   small amplitude limit
    & {\bf unstable} & {\bf stable} &
   {\bf stable}
   \\
   \hline
   Higher modes, $\sigma=-1$,
    general case
    & {\bf unstable}
    & {\em Hypothetically} {\bf unstable},
    if  $N$ exceeds some threshold. Otherwise {\bf stable}.
    & {\bf stable} for all $N$ which we have considered.\\
    \hline\hline
    \end{tabular}
    \caption{Comparison of the stability properties for the potentials
    $V(x)=x^2$, $V(x)=x^2+\kappa x^4$ (for $\kappa>0$) and $V(x)=x^4$. The results for $\kappa <0$ are not included in this table but discussed in the text}
    \label{Table3}
    \end{table}

\end{widetext}

\section{Conclusions and discussion}
\label{Conclusion}

Using a combination of different analytical and  numerical tools
including the analysis of the small amplitude limit, the nonlinear
WKB approximation, the Krein signature and direct numerical
simulations we have analyzed the stability properties of
higher-order nonlinear trapped modes for the GPE with different
potentials. First, we have reviewed the results for the harmonic
potential $V(x)=x^2$ and discussed how the stability of the modes
is essentially affected by the fact that levels are equidistant.
Next, we have  considered the weakly anharmonic potential
$V(x)=x^2+\kappa x^4$, $0<|\kappa|\ll 1$. Our results, summarized
in Table \ref{Table3}, lead to the conclusion that even a small
anharmonicity which does not affect essentially the shape of the
modes, improves drastically the stability properties of
higher-order modes due to the fact that none of these potentials
has an equidistant spectrum. We conjecture that the same situation
would take place also for more generic perturbation of harmonic
potential, for instance, by non-symmetric (e.g. cubic)
perturbation.

Then we have checked that in the case of stronger anharmonicity
$V(x)=x^4$ the stability/instability picture is similar to the case of
potential $V(x)=x^2+\kappa x^4$, $0<\kappa\ll1$, $\kappa>0$.
We have  studied the GPE with the potential $V(x)=x^6$ (the details
have not been discussed in this paper) and found that they reproduce
the same essential features.

It follows form the arguments presented, that the scenario for
appearance of instability induced by the equidistant spectrum of
the harmonic oscillator holds also for other classes of potentials
with equidistant spectra (for construction of such potentials see
\cite{Equidistant1,Equidistant2,Equidistant3}) or, more generally,
for potentials for which the spacing between some levels (not
necessarily adjacent) are equal.  In that situation, the splitting
of double eigenvalues for the operator ${\cal L}^2_n$ can lead to
complex eigenvalues in the linear stability problem.

An interesting point for further investigation,  is the effect of
the type of confining potential on the stability of higher order
modes in two spatial dimensions, e.g. the stability of vortices
under deformations of the potential. This subject has attracted a
lot of attention in the last years
\cite{VP1,VP2,VP3,VP4,Fin1,Michalache2006,Fin2,RadSymmKevrek2007,Watanabe}
and the methodology developed in this paper could be useful. In
fact, the situation is similar to the one considered above. In the
case of harmonic potentials the spectrum of corresponding
eigenvalue problem is equidistant; the corresponding
eigenfunctions are Gauss-Laguerre modes. Then, one can expect that
switching to anharmonic potentials can also change the stability
properties of vortices and other higher order modes.

Finally, we would like to mention another  practical implication of
the enhanced stability of nonlinear modes by the anharmonicity of the
trap potential. As it was suggested in~\cite{KevrekDyn} such modes can
grow from the eigenstates of the linear oscillator by increasing the
nonlinearity using Feshbach resonance management (in the language of
this paper this corresponds to the ``motion" along a nonlinear branch
starting from the bifurcation point as the number of particles
increases starting from zero). This fact can be used for the generation
of single solitons or even solitonic trains. The instability of the
nonlinear modes in the case of the harmonic potential was the major
obstacle for the practical implementation of that mechanism. However,
the idea becomes experimentally feasible if an anharmoic potential is
used since now higher order branches have a different stability and
thus can lead to stable solitons.

\acknowledgments

GA acknowledges the support from the President Program for Leading
Scientific Schools (Project 3826.2008.2.).  The work of VVK was
supported by the grant POCI/FIS/56237/2004 (European Program FEDER and
FCT, Portugal). VMPG is partially supported by grants FIS2006-04190 (Ministerio de Educaci\'on y Ciencia, Spain) and
PCI-08-0093 (Junta de Comunidades de Castilla-La Mancha, Spain).


\begin{thebibliography}{99}
\bibitem{Pitaev}
C. J. Pethick   and H. Smith  {\it Bose-Einstein Condensation in Dilute
Gases}   (Cambridge University Press,Cambridge, England, 2001);  L. P.
Pitaevskii, S. Stringari, Bose-Einstein condensation, Oxford (2003);
{\it Emergent Nonlinear Phenomena in Bose-Einstein Condensates Theory
and Experiment} Eds. P. G. Kevrekidis, D. J. Frantzeskakis, and R.
Carretero-Gonz\'alez (Springer, 2008).
\bibitem{Science1995}
     M. H.  Anderson, J. R. Ensher, M. R. Matthews, C. E. Wieman and
     E. A. Cornell,  Science, {\bf 269}, 198 (1995).
\bibitem{PRL1995}
     C. C. Bradley, C. A.Sackett, J. J. Tollett and R. G. Hulet,  Phys. Rev.
     Lett.  {\bf 75}, 1687 (1995)
\bibitem{EdwBur1995}
     M. Edwards and K. Burnett,   Phys. Rev. A, {\bf 51}, 1382 (1995);
     F. Dalfovo and S. Stringari,     Phys. Rev. A, {\bf 53}, 2477 (1996);
     P. A. Ruprecht, M. J. Holland, K. Burnett  and M. Edwards,  Phys. Rev. A, {\bf
     51}, 4704 (1995).
\bibitem{Yukalov1997}
     V. I. Yukalov, E. P. Yukalova, and V. S. Bagnato, Phys. Rev. A {\bf 56},
     4845 (1997);     V. I. Yukalov, E. P. Yukalova and V. S. Bagnato, Phys. Rev. A, {\bf 66} 043602 (2002);
     M. Brtka, A. Gammal and L. Tomio, Phys. Lett. A, {\bf 359}, 339
     (2006).
\bibitem{Kivshar2001}
     Yu. S. Kivshar, T. J. Alexander and S. K.Turitsyn, Phys. Lett. A, {\bf
     278}, 225 (2001).
\bibitem{MalomLasPhys2002}
     R. D'Agosta, B. A. Malomed and C. Presilla, Laser Physics, {\bf 12},
     37 (2002).
\bibitem{AgPres2002}
     R. D'Agosta  and C. Presilla, Phys. Rev. A, {\bf 65}, 043609 (2002).
\bibitem{KonKev} V. V. Konotop and P. G. Kevrekidis, Phys. Rev. Lett. {\bf 91}, 230402 (2003).
\bibitem{KevrekDyn}
     P. G. Kevrekidis, V. V. Konotop, A. Rodrigues, and D. J.
     Frantzeskakis, J. Phys. B: At. Mol. Opt. Phys. {\bf 38}, 1173 (2005).
\bibitem{AlfZez2007}
     G. Alfimov and D. Zezyulin, Nonlinearity, {\bf 20}, 2075 (2007).

\bibitem{PelKevr2007}      D. E. Pelinovsky and P. G. Kevrekidis, \texttt{arXiv.org/abs/cond-mat/0705.1016}
\bibitem{Carr2001}
     L. D. Carr, J. N. Kutz and W. P. Reinhardt, Phys.Rev. E, {\bf 63},
     066604 (2001).

\bibitem{cigar-shape} V. M. P\'erez-Garc\'{\i}a, H. Michinel, H. Herrero, Phys. Rev. A \textbf{57}, 3837 (1998).

\bibitem{VP1} J. J. Garc\'{\i}a-Ripoll, G. Molina-Terriza, V. M. P\'erez-Garc\'{\i}a, and
L. Torner, Phys. Rev. Lett. \textbf{87}, 140403  (2001).

\bibitem{VP2}  L.-C. Crasovan, G. Molina-Terriza, J. P. Torres, L. Torner, V. M. P\'erez-Garc\'{\i}a, D. Mihalache,
Phys. Rev. E \textbf{66}, 036612 (2002).

\bibitem{VP3}  L.-C. Crasovan,  V. Vekslerchik, V. M. P\'erez-Garc\'{\i}a, J. P. Torres, D. Mihalache, and L. Torner,
Phys. Rev. A {\bf 68}, 063609 (2003).

\bibitem{VP4}  L.-C. Crasovan, V. M. Perez-Garcia, I. Danaila, D. Mihalache,  Ll. Torner,
Phys. Rev. A \textbf{70}, 033605 (2004).

\bibitem{Fin1} M. Mottonen, S. M. M. Virtanen, T. Isoshima, and M. M. Salomaa, Phys. Rev. A \textbf{71}, 033626 (2005).

\bibitem{Michalache2006}
     D. Michalache, D. Mazilu, B. A. Malomed, F. Lederer,  Phys.Rev.A,
     {\bf 73}, 043615 (2006).

\bibitem{Fin2} V. Pietila,  M. Mottonen, T. Isoshima, J. A. M. Huhtamaki  and S. M. M. Virtanen,
Phys Rev. A \textbf{74}, 023603 (2006).

\bibitem{Watanabe} G. Watanabe, and C. J. Petick, \texttt{cond-mat/0701270}

\bibitem{RadSymmKevrek2007}
     G.Herring, L. D. Carr, R. Carretero-Gonz\'alez, P. G. Kevrekidis and D. J. Frantzeskakis.
    Phys. Rev. A, {\bf 77}, 023625 (2008)

\bibitem{AZKV} D. A. Zezyulin, G. L. Alfimov, V. V. Konotop, V. M. P\'erez-Garc\'{\i}a,
Phys. Rev. A \textbf{76}, 013621 (2007).

\bibitem{MacKay1987}
     R. S. MacKay, {\it Stability of equilibria of Hamiltonian systems},
     in {\it Hamiltonian Dynamical Systems},  R.S.MacKay and
     J.Meiss eds, Adam Hilger, 1987, pp. 137-153.

\bibitem{Gelfand}
     I. M. Gelfand. {\it Lectures on Linear Algebra}, Dover Publications (1989).

\bibitem{Kato}     T. Kato, \emph{Perturbation theory for linear operators}, Springer-Verlag,
     Berlin - Heidelberg -Ney-York (1966).
\bibitem{BendOrsz}
     C. M. Bender, S. A. Orszag. {\it Advanced mathematical methods for
     scientists and Engeneers}, McGraw-Hill book company, 1978.
\bibitem{GaussTail}
     M. Kunze, T. Kupper, V. K. Mezentsev, E.G.Shapiro and
     S.K Turitsyn,  Physica D, {\bf 128}, 273 (1999)
\bibitem{AS} {\it Handbook of Mathematical Functions}, M. Abramovitz and I. A. Stegun, eds,
     (National Bureau of Standards, 1972)
\bibitem{Szego} G. Szeg\"{o}, {\it Orthogonal polynomials}, Amer.Math.Soc,
      Colloquium publ, V. XXIII, Providence, Rhode Island (1939).
\bibitem{Equidistant1}
      V. M. Eleonsky and V. G. Korolev, J. Phys. A: Math. Gen., {\bf 28},
      4973 (1995)
\bibitem{Equidistant2}
      V. M. Eleonsky and V. G. Korolev, Phys. Rev. A, {\bf 55}, 2580,
      (1997)
\bibitem{Equidistant3}
      J. Morales, J. J. Pe\~{n}a, A. Rubio-Ponce, Theor. Chem.
      Accounts, {\bf 110}, 403 (2003).
\end{thebibliography}
\end{document}